\documentclass[aps,prab,reprint,superscriptaddress,nofootinbib]{revtex4-2}

\usepackage[T1]{fontenc}
\usepackage[utf8]{inputenc}
\usepackage{graphicx}
\usepackage{amsmath,amssymb}
\usepackage{bm}
\usepackage{comment}
\usepackage[colorlinks=true,linkcolor=blue,citecolor=blue,urlcolor=blue]{hyperref}
\usepackage{orcidlink}

\usepackage{xcolor} 

\begin{document}

	\title{Accessing the symmetric component of quadrupole misalignment through
		counter-rotating closed orbits in EDM storage rings}

	\author{I. Abedrabbo}
	\affiliation{Istinye University, Istanbul, T\"urkiye}
	\author{S. Hac{\i}\"omero\u{g}lu\,\orcidlink{0000-0002-8207-4219}}
    \email{selcuk.haciomeroglu@istinye.edu.tr}
	\affiliation{Istinye University, Istanbul, T\"urkiye}

	\begin{abstract}
		In the storage ring proposed to measure the proton electric dipole moment (EDM), quadrupole misalignments of a few microns mimic a genuine EDM through a geometric (Berry) phase; at a $10~\mu$m rms misalignment the false signal is up to a thousand times the target sensitivity of $10^{-29}~e\cdot$cm. The misalignment splits into an antisymmetric component, which distorts the closed orbit and is visible to the beam-position monitors, and a symmetric component, whose orbit signature is strongly suppressed. Standard single-beam orbit correction therefore removes only the antisymmetric part. The proposed design reaches the symmetric one by beam-based alignment (BBA), which is effective but relatively slow and invasive. Using symplectic beam and spin tracking, we show that correcting the two counter-rotating beams jointly recovers the symmetric component from the orbit itself — it is carried by the sum of the two orbits, where their difference is blind and a single beam far too weak to recover it. The static BPM offset is degenerate with the symmetric misalignment in this sum, so the scheme does not replace the initial absolute alignment. However the drift is differential and the offset cancels, so a continuous two-beam correction maintains the alignment after a single initial BBA calibration, reducing the repeated realignment that the original proposal of the proton EDM experiment requires.
	\end{abstract}
	\maketitle

	\section{INTRODUCTION}
	\label{sec:intro}
	\subsection{Frozen spin method and the false EDM}
	\label{sec:frozenspin}
	A permanent electric dipole moment violates both parity (P) and time-reversal (T) symmetries and, assuming CPT invariance, implies CP violation. Standard Model estimates $10^{-31} e\cdot$cm for proton EDM. Any measurement above indicates Physics Beyond Standard Model. The proposed storage ring proton EDM (pEDM) experiment~\cite{srEDM,omarov} relies on frozen spin method, where the protons are stored at a ``magic'' momentum ($p\approx 0.7 $ GeV/$c$). Then, the coupling between the EDM and the radial electric field of the deflectors precess the spin out of the plane, which is measured by polarimetry~\cite{polarimeter}, with speed
	\begin{equation}
		f \equiv \frac{dS_y}{dt}, \quad d=10^{-29} e\cdot\text{cm} \longleftrightarrow f \approx 1 \text{ nrad/s}.
	\end{equation}
	Certain couplings with magnetic dipole moment that can mimic this precession, called false EDM signal, must be eliminated. This paper investigates false EDM signals originating from misalignments of magnetic quadrupoles.

	The magnetic field at the center of a quadrupole magnet is zero. If it is misaligned, the particles experience magnetic fields proportional to the misalignment. A vertical misalignment $dy$ produces a radial magnetic field that precesses the spin in the vertical plane --- the same plane as the EDM signal. A horizontal misalignment $dx$ produces a vertical field that precesses the spin in the horizontal plane. These precessions are harmless alone because the spin precesses back and forth with these misalignments. However, if the quadrupoles are misaligned horizontally and vertically in any order, noncommutativity of the orthogonal precessions result in a residual out-of-plane precession. This is called the geometric (Berry) phase effect~\cite{berry1984,aharonov1987,farley2004,haciomeroglu2019}, and the false EDM is quadratically related to the RMS of the misalignments:
	\begin{equation}
		f_\text{false} \propto dx\cdot dy \quad \rightarrow \quad \vert f_\text{false} \vert \propto \sigma^2.
		\label{Eq:sigma_quadratic}
	\end{equation}
	As will be shown in Sec.~\ref{sec:benchmark}, our simulations produce the RHS as $\sigma^{2 \pm 0.01}$, consistent with Ref.~\cite{omarov}. A realistic misalignment of $\sigma = 10~\mu$m produces a false EDM signal, roughly 1000 times larger than the genuine one.

	\subsection{Ring design}
	\label{sec:ring}
	This study is made with the same symmetric-hybrid ring of Ref.~\cite{omarov}: composed of 24 FODO cells with electric  deflectors and magnetic quadrupoles. The focusing (QF) and defocusing (QD) quadrupoles have the same gradient ($g=0.2$ T/m) with opposite signs, separated by drift spaces and deflectors:
	\begin{equation*}
		\begin{split}
			&\text{QF} \rightarrow \text{Drift} \rightarrow \text{Deflector} \rightarrow \text{Drift} \rightarrow \\
			&\text{QD} \rightarrow \text{Drift} \rightarrow \text{Deflector} \rightarrow \text{Drift} \rightarrow \\
			&\text{QF...}
		\end{split}
	\end{equation*}
	The deflectors are basically cylindrical capacitors of 95.5 m radius. To store the beam at magic momentum, the electric field is set to $E_0=4.4$ MV/m. Some of the ring and beam parameters are listed in Table~\ref{tbl:ring_parameters}.
	\begin{table}[]
		\centering
		\caption{List of some ring and beam parameters, consistent with Ref.~\cite{omarov}.}
		\begin{tabular}{ll}
			\hline
			Parameter & Value \\
			\hline
			Number of FODO cells & 24 \\
			Number of quadrupoles & 48 \\
			Ring radius & 95.5 m \\
			Beam momentum & 0.7007 GeV/c \\
			Relativistic velocity & 0.598 \\
			Lorentz factor & 1.248 \\
			Betatron tunes ($Q_x, Q_y$) & (2.64, 2.3) \\
			\hline
		\end{tabular}
		\label{tbl:ring_parameters}
	\end{table}
	\subsection{Correction of quadrupole misalignment}
	\label{sec:correction}
	Ref.~\cite{omarov} proposes a two-component method for correcting the quadrupole misalignment. The first component is the beam based alignment (BBA), where the gradient of the quadrupole is slightly changed until a misaligned beam stops responding to the change. This method can be accelerated by AC excitation~\cite{marti} and be parallelized~\cite{huang}.
	The second component combines four configurations, namely clockwise (CW) and counter-clockwise (CCW) beams with quadrupole polarity switching ($g \rightarrow -g$). Subtracting the EDM measurements of the counter-rotating (CR) beams at reversed polarities eliminates the false EDM signal at ideal conditions (Eq. (C2) of Ref.~\cite{omarov}). This correction scheme is powerful, but not free. Polarity switching requires twice the measurement time. Also, the lattice must be highly symmetric between the two opposite polarities.

	We therefore investigate whether standard orbit correction (OC) methods can be useful. It typically relies on singular value decomposition (SVD) of the response matrix (R) and minimizing the closed orbit distortion. This approach is applied routinely since Chung, Decker and Evans~\cite{chung1993} and can operate continuously during data taking. The closed orbit is obtained by the beam position monitors (BPM) next to each quadrupole. On-line response-matrix refinement~\cite{ziemann}, symmetry correction~\cite{mirza} and orbit response modelling techniques~\cite{wegscheider} can be utilized as different treatments to the response matrix.

	It turns out that regardless of which technique is being used, simple orbit smoothing is not sufficient to eliminate this geometric phase effect. This is because the mechanisms for false EDM generation and closed orbit distortion are fundamentally different. However, OC with CR beams together is useful for suppressing the false EDM signal. This measurement still cannot replace the original proposal of Ref.~\cite{omarov} because BPM offset spoils it. Nevertheless, it can be useful for drift monitoring of the quadrupoles, once they are calibrated by BBA.

	\subsection{Contribution of this work}
	\label{sec:contribution}
	The central result of this work is that the symmetric component of quadrupole misalignment, largely invisible to conventional orbit correction, becomes observable through the summed closed orbit of counter-rotating beams, enabling continuous drift monitoring after an initial BBA calibration. The false EDM signal is not only defined by the RMS of the quadrupole misalignments, but also their pattern. Simple orbit smoothing can hardly correct the \textit{symmetric mode} of the misalignments, so it is not sufficient for the elimination of the false EDM. Closed orbits of the CR beams as summed reveal this symmetric mode, which allows continuous monitoring after first calibration. A short summary follows.

	Section~\ref{sec:mechanism} shows that the RMS of the quadrupole misalignments alone cannot represent geometric phase effect, yet a bilinear functional of the horizontal and vertical misalignment patterns does. Any misalignment pattern can be represented by two bases: symmetric and antisymmetric, indicating whether neighbouring quadrupoles are in the same or the opposite directions, respectively. Conventional orbit correction algorithms are almost blind to the symmetric component.

	The algorithm proposed in Ref.~\cite{omarov} (namely using BBA with CR beams), actually can eliminate this false EDM signal as long as the symmetry is conserved between opposite quadrupole polarities. More details are given in Sec.~\ref{sec:bba}. On the other hand, the sum of the closed orbit distortions of the CR beams reveals the symmetric component of the misalignments (Sec.~\ref{sec:croc}).

	The difficulty with this measurement is that, each closed orbit measurement suffers from the BPM offset, and their sum multiplies it by 2. Therefore, rather than making absolute measurements, this approach can be used to monitor the quadrupole drifts after the first BBA. Not requiring quadrupole polarity switch brings two advantages: 1) the measurements will be more tolerant to lattice errors, 2) no second storage (with flipped polarity) is needed for background rejection, hence the total experimental time is roughly halved. Perhaps BBA should still be done routinely, but the much faster and inexpensive OC can monitor the BBA period. In Sec.~\ref{sec:croc} we will discuss the operational details. Now, Section~\ref{sec:simulation} discusses the simulation tool for beam and spin dynamics.

	\section{SIMULATION METHOD}
	\label{sec:simulation}
	\subsection{Beam and spin tracking}
	\label{sec:tracking}
	All of the simulated results of this work were obtained by in-house beam and spin tracking codes. The implementation was benchmarked as described in Sec.~\ref{sec:benchmark}. They utilize Gauss-Legendre symplectic integrator~\cite{hairer} to estimate the \textit{one-particle} tracking inside the lattice as described in Sec.~\ref{sec:ring}. The simulations started with spin of the particle being in longitudinal direction. Then, the spin was precessed according to the Thomas-BMT~\cite{bmt1959} equations by the same integrator at every step.

	The genuine EDM as defined through the coupling parameter $\eta = 1.88 \times 10^{-15}$, corresponding to $d_p\approx 10^{-29}~e\cdot$cm, yielded $dS_y/dt = 9.81 \times 10^{-10}$ rad/s in the simulations. This value was reversed for the CR particle, consistent with Ref.~\cite{omarov}. All the false EDM estimations were made by artificially setting $\eta=0$, so that the false EDM signal was isolated. Gradient errors, quadrupole misalignments and tilts, and other studied systematic errors were introduced as ring parameters.

	\subsection{Analytical model of the response matrix}
	\label{sec:respmodel}
	We used an analytical response matrix to expose the mode structure of the closed-orbit response (Fig.~\ref{fig:fig_orbit_modes}). Its elements are given by the well known function
	\begin{equation}
		R_{ij} =  (KL)_j \frac{\sqrt{\beta_i \beta_j}}{2 \sin \pi Q}\cos(\vert \Delta \psi_{ij} \vert - \pi Q),
		\label{Eq:response_matrix}
	\end{equation}
	where $(KL)_j$ is the integrated gradient at the quadrupole $j$, $\beta_i$ and $\beta_j$ are the beta functions at the measurement and kick locations, $\Delta \psi_{ij}$ is the phase advance between the kick and measurement locations, and $Q$ is the betatron tune. When this analytical estimation was made with given lattice parameters (rather than estimated by simulations), it was consistent with the simulation results within a few percent for the vertical direction. However, for horizontal motion, the focusing effect of the deflectors introduces additional errors. Hence, we used the simulation-generated parameters for the estimation of the response matrix.

	\subsection{False EDM estimations}
	\label{sec:estimator}

	False EDM signal is the precession rate of the out-of-plane spin component: $dS_y/dt$. If an ideal particle is kicked by a few microns at the quadrupoles, it makes betatron oscillations at the same order, and sees a few microtesla oscillating field during storage. The coupling between this field and the magnetic dipole moment is many orders of magnitude larger than the EDM signal itself.

	We followed a two step algorithm to eliminate this background. First, we estimated the real closed orbit with the lattice errors. Then, for spin estimations, we injected the ideal particle in a way that it follows this real closed orbit. Secondly, we fit the vertical spin component to
	\begin{equation*}
		S_y(t) = a + bt + \sum_k [c_k \cos \omega_k t + d_k \sin \omega_k t],
	\end{equation*}
	and pick the secular slope $b$.

	\subsection{Benchmarking the estimations}
	\label{sec:benchmark}
	We benchmarked the estimator by following a single observable, the vertical spin precession rate, through three configurations. With the quadrupoles perfectly aligned and the genuine-EDM switch off, the estimator returned $\sim\!10^{-11}$~rad/s, consistent with zero at the numerical floor, so it fabricates no spurious signal. Switching on the EDM of $d = 10^{-29}~e \cdot$cm produces $dS_y/dt = 9.81\times10^{-10}$~rad/s, consistent with expected precession rate of Ref.~\cite{srEDM, omarov}. Finally, with the EDM switch off and the misalignments restored, the false signal grew quadratically with the RMS misalignment ($\vert f \vert \propto \sigma^p$, $p=2.00\pm0.01$ over $\sigma=2.5-10~\mu$m), and the corresponding beam separation fell in the $200-230~\mu$m range for a $10~\mu$m RMS misalignment, again consistent with Ref.~\cite{omarov}. Reversing the signs of the displacements flipped the signal as $f(+dx,+dy)=-f(+dx,-dy)=f(-dx,-dy)$, confirming its bilinear $dx\cdot dy$ origin (Eq.~\eqref{Eq:sigma_quadratic}).

	\section{FALSE EDM MECHANISM}
	\label{sec:mechanism}
	It is clear that the false EDM signal is proportional to the RMS of the quadrupole misalignments. However, there is an additional ingredient: the symmetric/antisymmetric components of the misalignment pattern. Quality of mapping between the geometric phase effect and the closed orbit requires a proper treatment of this structure.
	\subsection{Symmetric and antisymmetric components of misalignment}
	\label{sec:symanti}
	Let $v_x[2k]$ and $v_x[2k+1]$ be the misalignments of QF and QD respectively on the horizontal plane. These misalignments can be projected onto symmetric ($v_x^s$) and antisymmetric ($v_x^a$) basis as
	\begin{equation}
		\begin{split}
			v_x^s[2k] &= v_x^s[2k+1] = \frac{1}{2}\left( v_x[2k] + v_x[2k+1] \right), \\
			v_x^a[2k] &= -v_x^a[2k+1] = \frac{1}{2}\left( v_x[2k] - v_x[2k+1] \right).
		\end{split}
	\end{equation}
	which indicates that every quadrupole misalignment (QF or QD) can be decomposed into the symmetric and antisymmetric components of its cell:
	\begin{equation}
		v_x[n] = v_x^s[n] + v_x^a[n].
		\label{Eq:v_sym_anti}
	\end{equation}
	The same arguments hold for the vertical misalignments with $v_y, v_y^s$ and $v_y^a$.

	\begin{figure*}[!t]
		\centering
		\includegraphics[width=\linewidth]{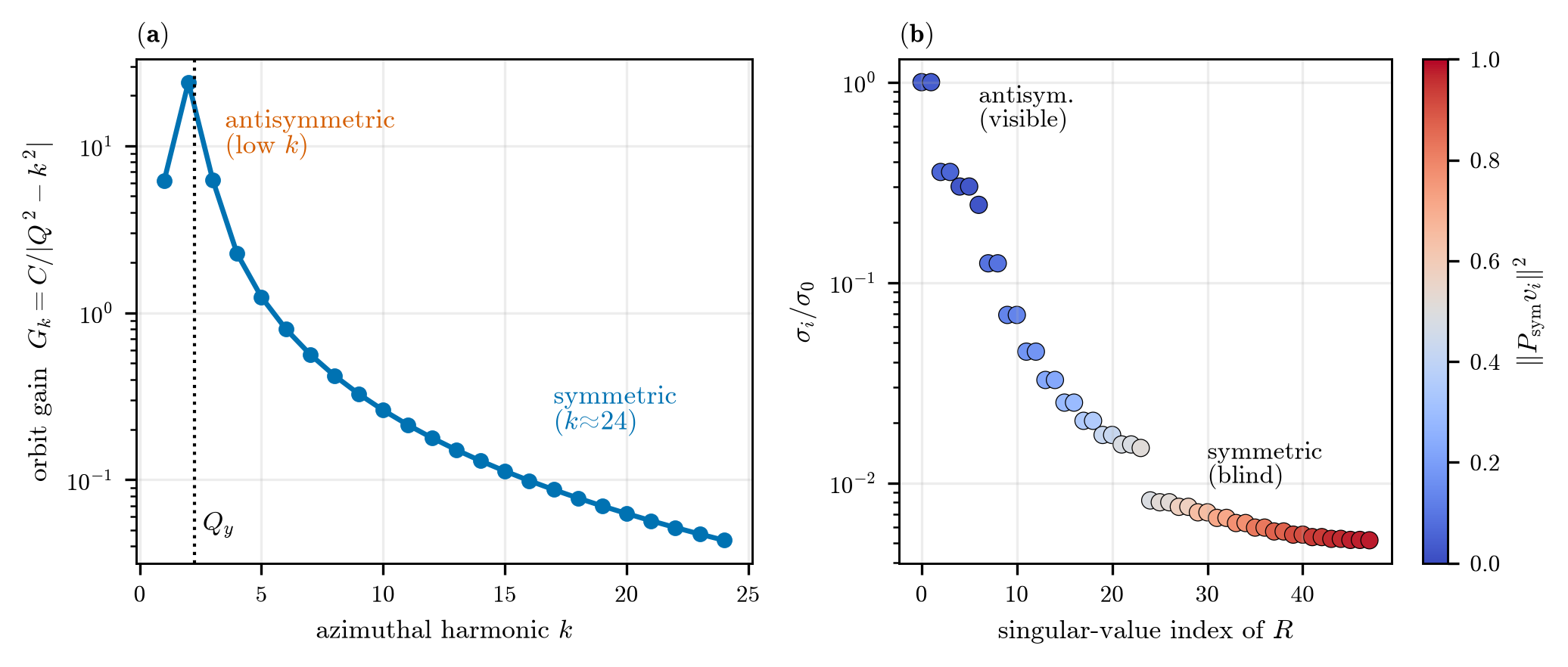}
		\caption{Visibility of misalignments on the orbit. (a) Orbit gain maximizes nearby the tune values ($k \approx Q$), and minimizes as $k$ goes further away from $Q$. (b) When the singular values of the SVD of a response matrix are ordered, the smallest ones correspond to the symmetric components.}
		\label{fig:fig_orbit_modes}
	\end{figure*}

	The antisymmetric misalignments kick the beam in the same direction and have long range effect over the ring. On the other hand, if the betatron tune is small compared to the number of FODO cells (See Table \ref{tbl:ring_parameters} for the proton EDM ring), the symmetric misalignments kick the beam in opposite directions, leaving a short range fingerprint on the closed orbit. But the long range effect of the kick is significantly cancelled.
	Rephrasing this statement, the $k^\text{th}$ Fourier component of the misalignment pattern affects the orbit through the gain parameter $G_k$ as
	\begin{equation}
		G_k \propto \frac{1}{\vert Q^2 - k^2 \vert}.
		\label{Eq:orbit_gain}
	\end{equation}
	The symmetric modes represent the contributions from $k=24$ for the proton EDM ring. They are strongly suppressed because $Q^2 \ll k^2$. As an example, $k=1$ component is around 100 times larger than the $k=24$ component. This behavior is summarized in Fig.~\ref{fig:fig_orbit_modes}. However, the effect of the very same field on the spin is not negligible, because of the geometric phase mechanism. While faintly visible to the orbit, this effect is quite strong on the spin. In conclusion, standard orbit correction methods have limited power against the geometric phase effect.

	\subsection{False EDM in terms of bilinear functional}
	\label{sec:bilinear}
\begin{figure*}[!t]
		\includegraphics[width=\linewidth]{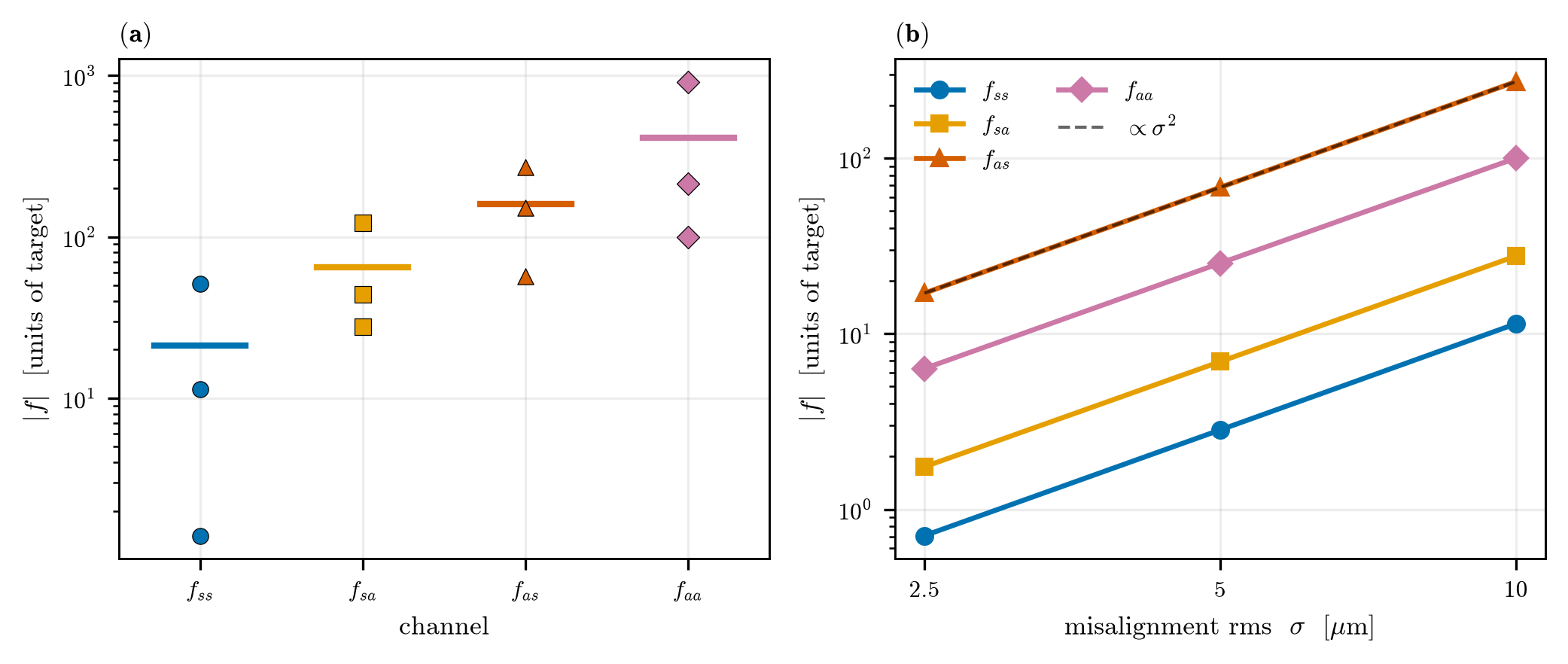}
		\caption{Four channels of the false EDM. The data was obtained by first misaligning the quadrupoles with symmetric and antisymmetric configurations. (a) The simulations were made with three seeds. Similar to the orbit response, the false EDM signal also becomes larger as the antisymmetric mode dominates.  (b) Every channel independently obeys the $\sigma^2$ scaling. This data is obtained with one seed. }
		\label{Fig:channels}
	\end{figure*}

	Following the above conventions, Eq.~\eqref{Eq:sigma_quadratic} can be written for every FODO cell as
	\begin{equation}
		f =  \sum_i \sum_j W_{ij} v_{x,i}v_{y,j} = F(v_x, v_y),
	\end{equation}
	where $i,j$ count the neighbor quadrupoles, $W_{ij}$ is the weight for each FODO cell, defined by the lattice. It is worth noting that $v_{x,i}$ and $v_{y,j}$ belong to two neighbor quadrupoles, not one.
	Splitting both of them into symmetric and antisymmetric components (Eq.~\eqref{Eq:v_sym_anti}), one gets
	\begin{equation}
		f= \underbrace{F(v_x^s, v_y^s)}_{f_{ss}} +
		\underbrace{F(v_x^s, v_y^a)}_{f_{sa}} +
		\underbrace{F(v_x^a, v_y^s)}_{f_{as}} +
		\underbrace{F(v_x^a, v_y^a)}_{f_{aa}}.
	\end{equation}
	Figure~\ref{Fig:channels} investigates this mixing. Similar to Fig.~\ref{fig:fig_orbit_modes} of the orbit response, the antisymmetric content ($f_{aa}$) is more dominant than the symmetric one ($f_{ss}$). However, this dominance is not as strong as the orbit case. Therefore, standard orbit correction suppresses the RMS misalignment $\sigma$, but does not guarantee the elimination of the geometric phase effect. The experimental target ($d_p=10^{-29}~e \cdot$cm) of the proton EDM experiment requires correcting the symmetric component.

	\section{APPLICATION OF BEAM BASED ALIGNMENT}
	\label{sec:bba}
	Beam based alignment (BBA) does not suffer from the above mentioned imbalance between spin and orbit, because it relies on correcting the quadrupoles individually, rather than relying on the global closed-orbit response. Consequently, the symmetric and antisymmetric structure of the misalignment pattern does not affect the alignment procedure. For this reason, the proposal of Ref.~\cite{omarov}, BBA with CR beams followed by quadrupole polarity switch significantly suppresses the geometric phase effect.

	It is worth noting that BBA does not require inverting the response matrix. Therefore, it is more immune to the discrepancy between the symmetric and antisymmetric modes. It becomes a significant difficulty for SVD algorithms (See Fig.~\ref{fig:fig_orbit_modes}).

	Application of BBA requires iterative minimization of the dipole kick of the quadrupole magnet. The kick becomes zero when the beam passes through the magnetic center of the quadrupole. BBA aims to minimize this kick by correction dipoles next to the quadrupole. The quadrupole gradient is slightly varied in order to isolate the signal from the background sources, making the measurement largely insensitive to BPM offset and slowly varying effects.

	\subsection{BBA with realistic optics}
	\label{sec:bba_realistic}

\begin{figure}
		\centering
		\includegraphics[width=\linewidth]{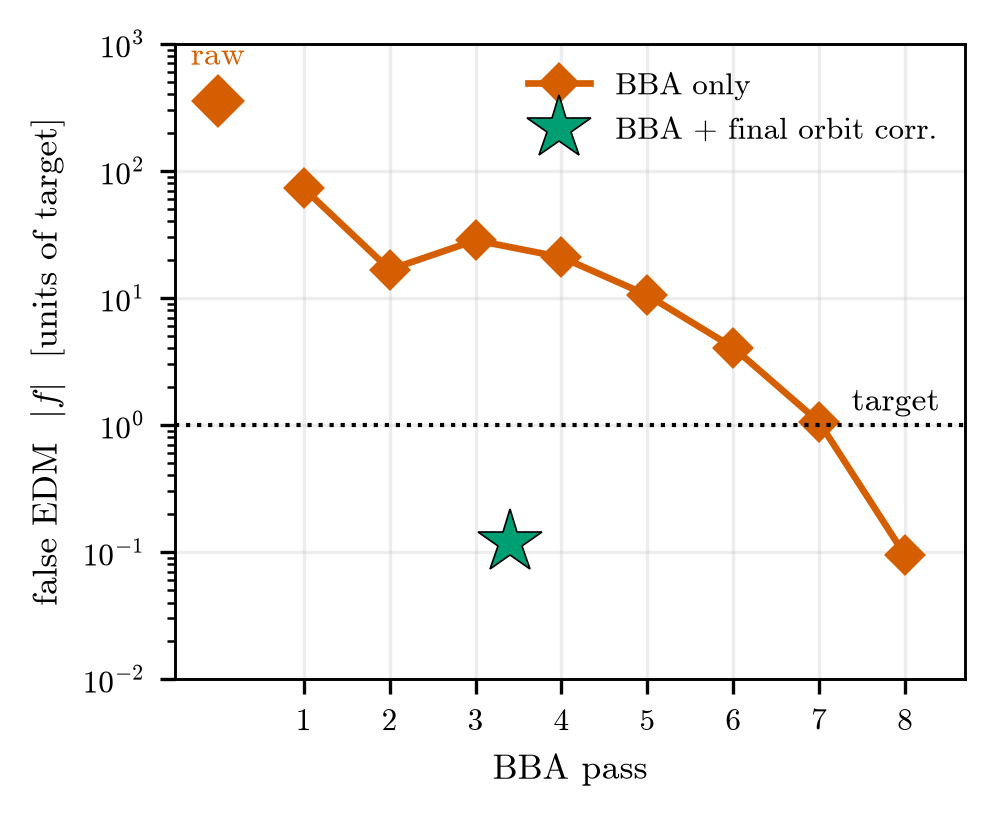}
		\caption{BBA algorithm requires several iterations to reduce the misalignments compatible with the EDM target. The false EDM signal is given at top-left before the operations. Applying conventional OC after a few iterations is effective for faster correction. The simulations were made with $1\%$ RMS gradient errors. At each iteration, the quadrupoles were aligned one by one.}
		\label{fig:bba_convergence}
	\end{figure}

	As shown in Ref.~\cite{omarov}, the BBA algorithm with CR beams and quadrupole polarity switch reduces the false EDM signal to the target value in ideal conditions. Our simulations with $\approx 1\%$ random gradient error yielded a \textit{breathing} effect (see Appendix \ref{app:breathing}). Variation of the gradient of a quadrupole shifts the focus-related parameters of the \textit{whole} ring slightly, especially the betatron tunes. Therefore, the response of all other quadrupoles, hence the BPM measurements also get shifted. The modulation does not help isolating this effect, because it is the source. However, this systematic error is proportional to the COD, and can be minimized by iteratively applying BBA. As Fig.~\ref{fig:bba_convergence} shows, the false EDM reduced to the target value after eight iterations.

	Unlike the vertical ones, the horizontal misalignments could not be corrected well with analytical response matrix, where the electrical deflectors were also treated as drifts and the resultant response matrix was deviated on the horizontal plane. Therefore, we constructed the response matrix by simulation, and applied the BBA iterations afterwards.

	Since BBA does not distinguish the symmetric and antisymmetric components, it reduces them simultaneously. This opens up an opportunity to use OC after enough reduction of the symmetric components by BBA. The green star in Fig.~\ref{fig:bba_convergence} shows that applying OC after the third BBA iteration performs equivalently with eight iteration BBA. However, we will show in Sec.~\ref{sec:croc} that OC can be utilized more effectively to save the experimental time. It relies on CR beam measurements too.

	\begin{figure}
		\centering
		\includegraphics[width=\linewidth]{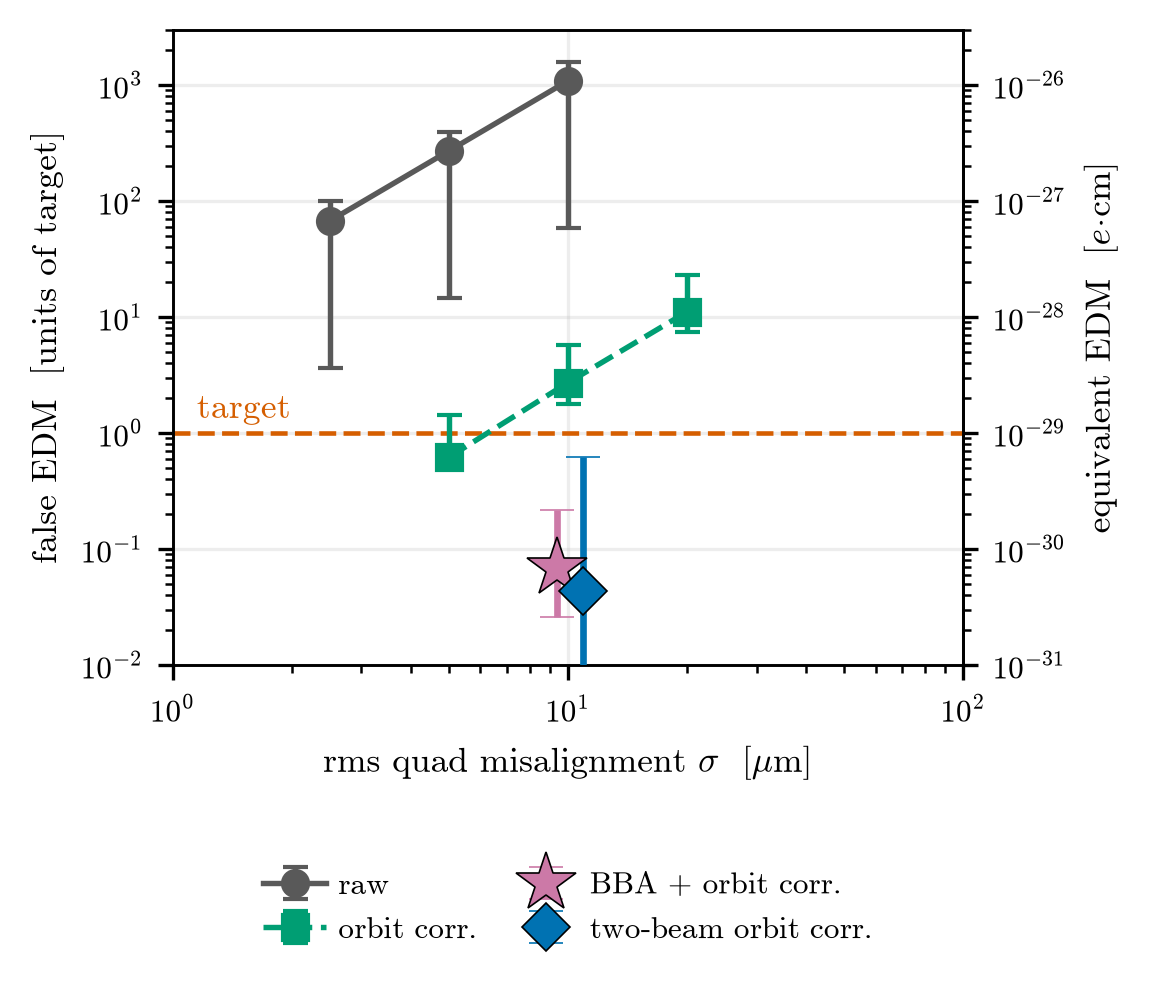}
		\caption{Summary of the tested methods for geometric phase suppression. The circles show the false EDM signal without any correction. The squares show the best results with OC alone. The star belongs to BBA after 8 iterations (or OC after 3 BBA iterations). The diamond shows that OC with CR beams performs similarly to BBA. Each test was made over 5 seeds.}
		\label{fig:orbit_suppression}
	\end{figure}

	\section{ORBIT CORRECTION WITH COUNTER-ROTATING BEAMS}
	\label{sec:croc}
	We showed that the standard orbit correction approach has limited access to the symmetric component, and fails at suppressing the geometric phase effect to the experimental target. The failure occurred when realistic errors (like BPM offsets of order $100 ~\mu$m, BPM white noise of order $1~\mu$m, gradient errors of order $1\%$, quadrupole tilts) were included in the simulations.

	Varying the gradient in fact can eliminate the BPM offset because it is the common background regardless of the lattice parameters. This can be applied by modulating the quadrupole gradient as well. The difficulty in this approach is about the inversion of the response matrix.
	The orbit difference for different gradients is determined by the difference response matrix $\Delta R$ as
	\begin{equation}
		\begin{split}
			\Delta x &= \Delta R v_x, \\
			\Delta y &= \Delta R v_y,
		\end{split}
	\end{equation}
	where $\Delta x$ and $\Delta y$ are the BPM measurement of the COD. It can be shown that if the gradient is changed by $\epsilon_g$, the inverse of the difference scales as
	\[\vert \vert \Delta R^{-1} \vert \vert \approx \vert \vert R^{-1} \vert \vert / \epsilon_g. \]
	In other words, the weakness of this method is the conditioning: $3.7 \times 10^{4}$ for the pEDM lattice. As the smaller singular values belong to the symmetric components (Fig.~\ref{fig:fig_orbit_modes}), the matrix inversion selectively underperforms at symmetric components. Attempts like Tikhonov regularization~\cite{tikhonov1977} did not help either, because they cut off the lower singular values -- the symmetric modes. Even with 0.1\% RMS gradient, the symmetric components were reconstructed with $\approx 250\%$ error (See Appendix \ref{app:orb_diff_inv}). This limitation cannot be overcome by BPM sensitivity, because the noise is coherent (not white).

	Figure~\ref{fig:orbit_suppression} shows how different approaches perform for this task. As the blue diamond at that plot shows, orbit correction can in fact be effective if CR beams are used. We will discuss it in the next section.

	\subsection{Extracting the symmetric component with counter-rotating beams}
	\label{sec:extract}
\begin{figure*}[tb]
		\centering
		\includegraphics[width=1\linewidth]{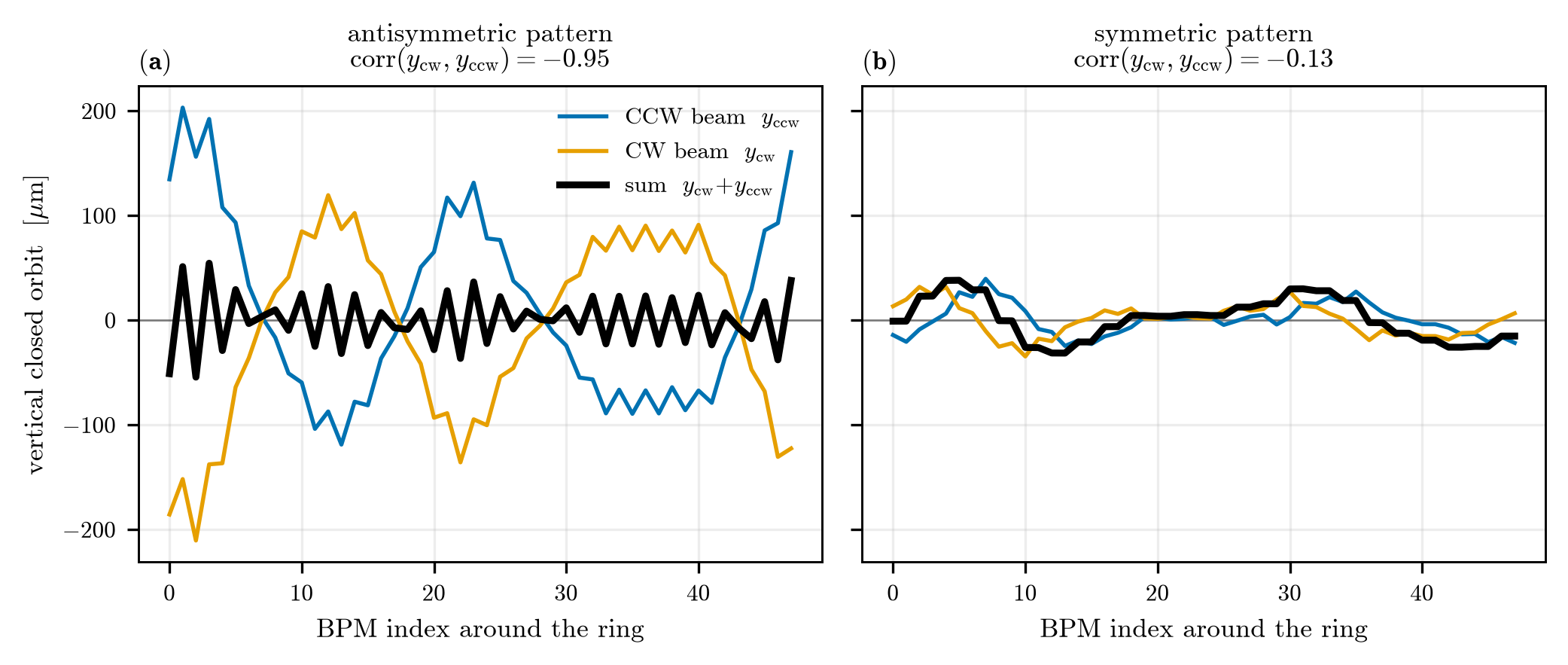}
		\caption{Closed orbit distortion (COD) for a $10~\mu$m rms misalignment. (a) For an antisymmetric misalignment the two counter-rotating orbits are almost opposite, so their sum is a tiny residual (black). (b) For a symmetric misalignment the two orbits are almost the same up to a small phase shift, so their sum (black) survives. The resulting residual—the black sum—is of the same magnitude for the symmetric and antisymmetric components, unlike the individual orbits.}
		\label{fig:orbit_cw_ccw}
	\end{figure*}

	The quadrupole misalignments in a FODO cell exert opposite forces ($\bm F=q~\bm v \times \bm B$) on the counter-rotating beams. Hence, the antisymmetric components of the closed orbit are almost opposite [Fig.~\ref{fig:orbit_cw_ccw}(a)]. It is not perfect though, because the CR beams pass through the quadrupoles in the opposite order, leading to a residual distortion (black curve). For the symmetric component the orbit depends on this order, so reversing both the force and the order leaves nearly the same pattern with a phase shift [Fig.~\ref{fig:orbit_cw_ccw}(b)]. Hence, after summation, the residual COD for the symmetric and antisymmetric patterns are of the same order, as opposed to the one-beam case.

	The two counter-rotating beams circulate together and are read on
	the same BPMs (separated by bunch timing), so both closed orbits are known at every
	monitor. Both symmetric and antisymmetric components can be recovered by stacking the two response matrices and solving a single least-squares correction. More details are given in Appendix \ref{app:sum}.

	On the raw $10~\mu$m machine, without beam-based alignment, two-beam correction drives the symmetric residual below the genuine-EDM target (down to $0.1$--$0.7~\mu$m). A $0.2\%$ rms quadrupole gradient error together with a $0.2$~mrad quad tilt keeps it well under target, and only $5\%$ $\beta$-beat brings it close.

	Separation between the counter beams eliminates the effect of the BPM offset; it doubles when summed. So, the BPM offset is the major limitation preventing it from being a standalone method. But it can be used as a drift monitor after BBA, since the BPM offset is almost fixed in time.

	A major advantage of this method is statistics. In the reference scheme, the quadrupole polarities must be flipped for additional cancellation of the common mode background. Our proposed orbit correction scheme does not require that quadrupole polarity flip. So, the same sensitivity can be achieved in one run, rather than two.

	\subsection{Drift and the time budget}
	\label{sec:drift}
	The scheme is meant to run continuously, as a drift monitor rather than a one-shot alignment. Because each BPM sits next to its quadrupole, the two are likely to exhibit correlated mechanical motion~\cite{rossbach}, so the two-beam sum tracks this common motion and the accumulated correction is a running measure of how far the machine has drifted since the last calibration. Beam-based alignment then only sets the absolute baseline and is renewed when this drift crosses the sub-target threshold---turning an expensive absolute measurement into an occasional recalibration, with the cheap two-beam orbit covering the interval between.

	\section{CONCLUSIONS}
	\label{sec:conclusion}

	Our investigations regarding the geometric phase effect due to quadrupole misalignments in the proton EDM ring revealed that the false EDM signal is a bilinear functional of the misalignment pattern. This pattern is composed of symmetric and antisymmetric components. The latter is dominant in both orbit and the spin (false EDM) measurements. The former is hardly visible in orbit measurements, while its effect on the false EDM signal is not negligible.

	The proposal of Ref.~\cite{omarov}, BBA with CR beams, followed by quadrupole polarity switch at consecutive runs is robust in suppressing the both components, hence the false EDM signal. The symmetric mode is buried under noise and other realistic systematic effects in conventional orbit correction methods. However, summing the closed orbits of the counter-rotating beams provides access to the symmetric component, eventually leading to a faster suppression of the orbit and the false EDM. It brings an additional advantage -- no requirement for quadrupole polarity switch. Hence, by removing the quadrupole-polarity reversal required in the reference scheme, it has the potential to reduce the required running time by approximately a factor of two.

	Rather than replacing beam-based alignment, our proposed method complements it by providing a practical means for continuous drift monitoring between BBA calibrations.

	\begin{acknowledgments}

	This work was supported by the Scientific and Technological Research Council of Türkiye (TÜBİTAK) under the 1001 Program (Project No.~123F256).

	Generative AI tool (Claude 4 Opus) was used to assist in software development, debugging, and figure-generation scripts. All scientific methodology, validation, interpretation, and conclusions were performed and verified by the authors.
\end{acknowledgments}

	\appendix

	\section{Optics breathing}
	\label{app:breathing}

\begin{figure}
		\centering
		\includegraphics[width=1\linewidth]{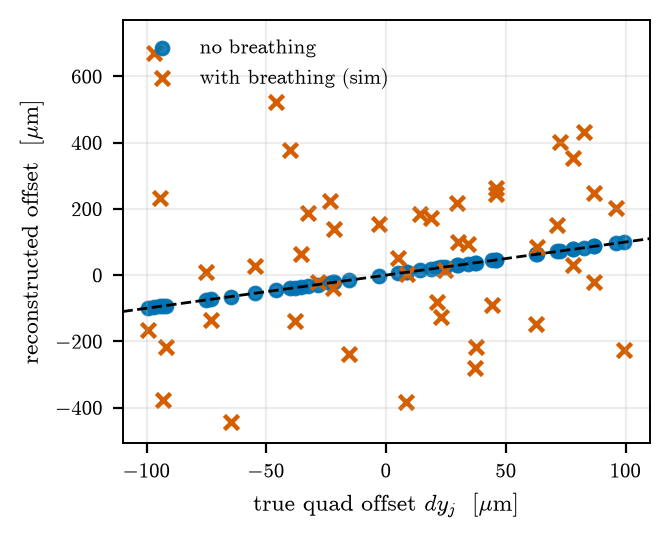}
		\caption{Reconstruction of misalignments when quadrupoles are modulated. If the breathing effect is artificially cancelled, the reconstruction works well. But the breathing buries the signal into noise.}
		\label{fig:breathing}
	\end{figure}

	Optics breathing effect is the second big obstacle in orbit difference approach, if the quadrupoles are modulated. Consider a change in gradient of a single quadrupole. It causes a slight change in betatron-related parameters. Then, each response matrix element (Eq.~\eqref{Eq:response_matrix}) will slightly change, introducing error in the reconstruction. Especially if all the quadrupoles are modulated at different frequencies, the response matrix elements have large contribution from these errors. Simulation results are shown in Fig.~\ref{fig:breathing}.

	\section{Inversion of orbit difference}
	\label{app:orb_diff_inv}
\begin{figure}
		\centering
		\includegraphics[width=1\linewidth]{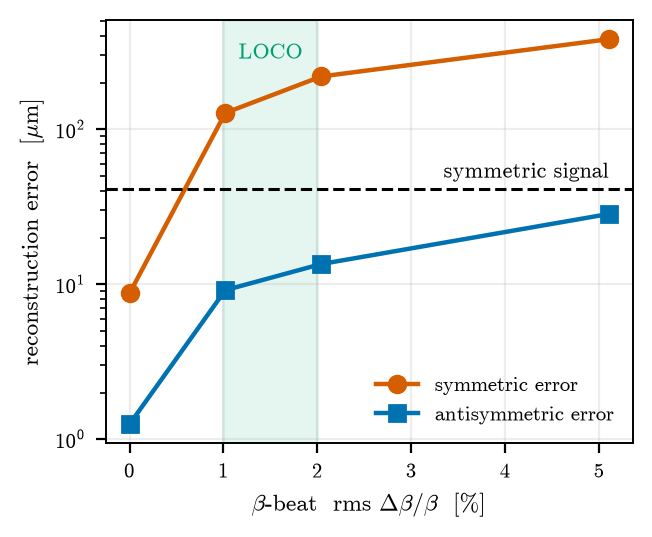}
		\caption{Reconstruction error of the orbit difference approach as a function of the beta beat. Error in the symmetric component becomes roughly 15 times larger than the antisymmetric component.}
		\label{fig:orbit_lock}
	\end{figure}

	Orbit difference cancels the static offsets, but the resulting difference response matrix $\Delta R$ has a high conditioning number, at the order of $10^4$. Hence, the method becomes extremely sensitive to errors. As shown in Fig.~\ref{fig:orbit_lock}, the symmetric error quickly gets larger than the signal with beta beat. Regularizations like Tikhonov does not make it better, because they cut off the symmetric component.

	\section{Access to the symmetric component with the counter-rotating sum}
	\label{app:sum}

	A two-dimensional toy model shows what the two counter-rotating orbits carry, how the sum is measured, and how the misalignment is read back from it. In one plane write the
	misalignment as $v=(a,\,s)^{\!\top}$, an antisymmetric component $a$ and a symmetric
	component $s$ (in the full ring each stands for a $24$-dimensional family; here they are
	single coordinates). An antisymmetric component drives a large closed orbit, a symmetric
	one only a small orbit, at different harmonics, so the one-beam response is diagonal,
	$R_{\rm CW}=\mathrm{diag}(G,g)$ with $G\gg g$. Beam reversal flips the antisymmetric orbit but not the symmetric one --up to a small phase shift-- so $R_{\rm CCW} = \mathrm{diag}(-G,g)$, and $y_{\rm CW}=R_{\rm CW}v=(Ga,\,gs)^{\!\top}$, $y_{\rm CCW}=R_{\rm CCW}v=(-Ga,\,gs)^{\!\top}$.

	The two beams are stored together and read on the same BPMs, told apart by bunch timing,
	so each monitor delivers both orbits; every reading also carries the monitor's static
	offset $b=(b_a,\,b_s)^{\!\top}$, common to the two beams. Writing $R_{\rm sum}=R_{\rm CW}+R_{\rm CCW}=\mathrm{diag}(0,2g)$ and $R_{\rm CW}-R_{\rm CCW}=\mathrm{diag}(2G,0)$, the difference and the offset-carrying sum are
	\begin{equation}
		\begin{split}
			y_{\rm CW}-y_{\rm CCW} &= (R_{\rm CW}-R_{\rm CCW})\,v = \begin{pmatrix}2Ga\\[2pt] 0\end{pmatrix},\\[4pt]
			y_{\rm CW}+y_{\rm CCW} &= R_{\rm sum}\,v + 2b = \begin{pmatrix}2b_a\\[2pt] 2gs+2b_s\end{pmatrix}.
		\end{split}
		\label{eq:sumdiff}
	\end{equation}
	$R_{\rm sum}$ is itself a $2\times2$ matrix, and a singular one: its antisymmetric entry is zero, so the sum responds only to $s$.

	The misalignment is read back one component at a time. From the difference,
	\begin{equation}
		a=\frac{y_{\rm CW}-y_{\rm CCW}}{2G},
	\end{equation}
	free of the offset (it cancelled) and well conditioned. From the sum,
	\begin{equation}
		s=\frac{(y_{\rm CW}+y_{\rm CCW})-2b_s}{2g},
	\end{equation}
	which needs the small gain $g$ inverted --the ill-conditioning of the symmetric channel-- and, worse, the offset $2b_s$: with $b\sim100~\mu$m$\,\gg gs$, a static symmetric
	misalignment and a static offset are degenerate in the sum and cannot be separated by a
	single measurement. What removes $b_s$ is time: it is static, so the drift of the sum gives $\Delta s=\Delta(y_{\rm CW}+y_{\rm CCW})/2g$ with the offset gone. The scheme is thus a symmetric-drift monitor; the absolute symmetric level still needs beam-based alignment (Sec.~\ref{sec:croc}).

	These single divisions are the toy's stand-in for a fit. In the ring, $v$ has one entry
	per quadrupole and each orbit is sampled at every BPM, so the two beams give twice as many equations as unknowns; $a$ and $s$ are then recovered not by division but by the
	least-squares solution
	\begin{equation}
		\hat v = S^{+}\begin{pmatrix}y_{\rm CW}\\ y_{\rm CCW}\end{pmatrix},
		\qquad
		S=\begin{pmatrix}R_{\rm CW}\\ R_{\rm CCW}\end{pmatrix},
	\end{equation}
	with $S^{+}$ the regularized pseudo-inverse.

	Two points are worth noting. The parity is approximate --the beams cross each cell in the opposite order-- so the antisymmetric orbit does not cancel exactly in the sum but leaves the small residual seen as the black curve of Fig.~\ref{fig:orbit_cw_ccw}(a); the symmetric channel in practice is that measured residual, equal to $2gs$ up to the phase shift. And $g$ here stands for a whole family of symmetric gains.

\end{document}